\documentclass[12pt,preprint]{aastex}

\shorttitle{Optical Monitoring of S5~0716+714}
\shortauthors{Wu et al.}

\received{}
\accepted{}

\begin{document}

\title{Optical Monitoring of BL Lacertae Object S5~0716+714 with a Novel
Multi-Peak Interference Filter}

\author{Jianghua Wu, Xu Zhou, Jun Ma, Zhenyu Wu, Zhaoji Jiang, Jiansheng Chen}
 \affil{National Astronomical Observatories, Chinese Academy of Sciences,
        20A Datun Road, Beijing 100012, China}
 \email{jhwu@bao.ac.cn}

\begin{abstract}
We at first introduce a novel photometric system, which consists of a Schmidt
telescope, an objective prism, a CCD camera, and, especially, a multi-peak
interference filter. The multi-peak interference filter enables light in
multi passbands to pass through it simultaneously. The light in different
passbands is differentially refracted by the objective prism and is focused
on the CCD separately, so we have multi ``images" for each object on the CCD
frames. This system enables us to monitor blazars {\sl exactly} simultaneously
in multi wavebands on a single telescope, and to accurately trace the color
change during the variation. We used this novel system to monitor the BL
Lacertae object S5~0716+714 during 2006 January and February and achieved a
very high temporal resolution. The object was very bright and very active
during this period. Two strong flares were observed, with variation amplitudes
of about 0.8 and 0.6 mags in the $V'$ band, respectively. Strong
bluer-when-brighter correlations were found for both internight and intranight
variations. No apparent time lag was observed between the $V'$- and $R'$-band
variations, and the observed bluer-when-brighter chromatism may be
mainly attributed to the larger variation amplitude at shorter wavelength.
In addition to the bluer-when-brighter trend, the object also showed a bluer
color when it was more active. The observed variability and its color behaviors
are consistent with the shock-in-jet model.
\end{abstract}


\section{INTRODUCTION}
Among the family of active galactic nuclei (AGNs), blazars manifest their
prominent characteristic as rapid and strong variability nearly throughout the
entire electromagnetic wave bands. In the unified scheme of AGNs, blazars are
those objects with their relativistic jets pointed basically to the observers
\citep[e.g.,][]{urry95}. The jet is believed to originate from and be
accelerated by a rotating supermassive black hole surrounded by an accretion
disk. There are two types of blazars. Depending on whether or not they show
strong emission lines in their spectra, they are called flat-spectrum radio
quasars (FSRQs) or BL Lac objects.

Several mechanisms, still in competition and modification, have been proposed
to explain the variability of blazars \citep[see a review by][]{wagner95}.
The most common suggestion is the shock-in-jet model \citep[e.g.,][]
{marscher85,qian91}, in which shocks form at the base of the jet and propagate
downstream, accelerating electrons and compressing magnetic fields, and
resulting in the observed variability. Alternative but less accepted scenarios
include the interstellar scintillation (ISS) \citep[e.g.,][]{fiedler87,
rickett01}, the microlensing effects \citep{nottale86,schneider87}, the
instability in accretion disk \citep[e.g.,][]{abramowicz91,chak93,mangalam93},
and the geometrical effects \citep[e.g.,][]{camen92}. Some authors even
suggested a combination among them \citep[e.g.,][]{wu05,dolcini05}.
For some special blazars, for instance, the BL Lac object OJ~287, which is the
only blazar that shows convincing evidences for a 12-year period in its
optical variability \citep{sillanpaa88}, binary black hole model was
introduced to explain the observed periodicity \citep[e.g.,][]
{sillanpaa88,valtaoja00,liu02}. In brief, for the
basic question in blazar studies, i.e., the variation mechanism, there
exist various scenarios that still need to be verified and/or modified by
future elaborate observations.

In fact, an important but simple factor can discriminate among these various
scenarios. It is the spectral index or color. The shock-in-jet model predicts
spectral changes during the variations \citep[e.g.,][]{kirk98}, whereas the
geometrical or light house effects do not. The ISS is frequency-dependent, but
the micro-lensing effects are not. Moreover, these different mechanisms lead
to different timescales of variability and different profiles of light curves.
For example, the shock-in-jet and ISS will result in irregular light curves,
the geometrical effects are usually linked to periodic variations, while
micro-lensing effects should produce strictly symmetric light curves. By
combining the spectral index, timescale, and light curve profile, the
variation mechanism can be constrained.

Therefore, in order to constrain the variability mechanisms in blazars, the
crucial is to obtain the accurate
spectral index. The spectral index is calculated by using two photometric
measurements in two different wavelengths. The traditional blazar monitoring
on one telescope is to make observations in two or more wavelengths one by
one, i.e., one can make an exposure with filter A at first, then change to
filter B and make the second exposure. After that one again change to filter
C and make the third exposure, etc. The filters are usually used in a cyclic
pattern. With this approach, one can achieve {\sl quasi}-simultaneous
observations in two or more wavelengths on a telescope. If the brightness of
the target and the status of the instruments are both stable, one can
reasonably calculate the spectral index with these quasi-simultaneous
photometries.
For blazars, however, the case is very different. Blazars are highly variable,
with timescale as short as hours or even minutes. When the exposure with
filter A is completed and a new exposure with filter B begins, the intrinsic
brightness of the blazar may have changed. It will naturally result in large
error or completely wrong result when calculating the spectral index by using
the results of these two exposures. Therefore, all previous investigations
related to spectral index or color of those highly active blazars, if based on
quasi-simultaneous multi-waveband monitoring, have relatively low confidence
level.

In this paper, we introduce a novel photometric system, which includes a
Schmidt telescope, an objective prism, a CCD camera, and a multi-peak
interference filter. This new system enables us to make photometries {\sl
exactly} simultaneously in multi wavebands with a single exposure on a
telescope, thus enables us to obtain the {\sl exact} spectral index of blazars
during the variations. The simultaneous exposures in multi wavebands also
have the advantage of being invulnerable to the possible instability of
instruments and weather.

The BL Lac object S5~0716+714 is chosen as the first target of
this new photometric system. It is very bright among all BL Lac objects, and
it has a duty cycle of 1 \citep[e.g.,][]{wagner95}, which means it is almost
always active. In their monitorings of S5~0716+714, \citet{villata00} have
determined a steepest recurrent variation slope as 0.002 $\rm{mag\,min^{-1}}$
on both rising and decreasing phases, and \citet{nesci02} found typical
variation rate of 0.02 $\rm{mag\,hr^{-1}}$ and a maximum rising rate of 0.16
$\rm{mag\,hr^{-1}}$. A maximum rising rate of 0.1 $\rm{mag\,hr^{-1}}$ was also
reported by \citet{wu05}. Most recently, \citet{montagni06} claimed a highest
rate of magnitude variation of $0.10-0.12\,\rm{mag\,h^{-1}}$. So this object
with rapid and large-amplitude
variation is very suitable to be monitored with our new photometric system.

We monitored S5~0716+714 with this system in 2006 January and February. It
is, to our knowledge, the first time that a multi-peak interference filter
be used in blazar monitoring. Here we present the results, and
based on accurate color tracing during the variation, we discuss the variation
mechanism of this object.

\section{ObSERVATIONS AND DATA REDUCTIONS}

\subsection{The Photometric System}

The new photometric system consists of a Schmidt telescope, an objective
prism, a CCD camera, and a multi-peak interference filter. The Schmidt
telescope is 60/90 cm in diameter and is located at the Xinglong Station of
the National Astronomical Observatories of China (NAOC). The CCD is a thick
CCD, so it is
not sensitive in blue wavelengths. It has a pixel size of $15\,\micron$ and
a field of view of $58\arcmin\times58\arcmin$, resulting in a resolution of
$1\farcs7\,\rm{pixel}^{-1}$. When this system is used for blazar monitoring,
we read out only the central $512\times512$ pixels or
$14\farcm5\times14\farcm5$, which is adequate to cover the target blazar
and the reference stars around it.

\begin{figure}
\plotone{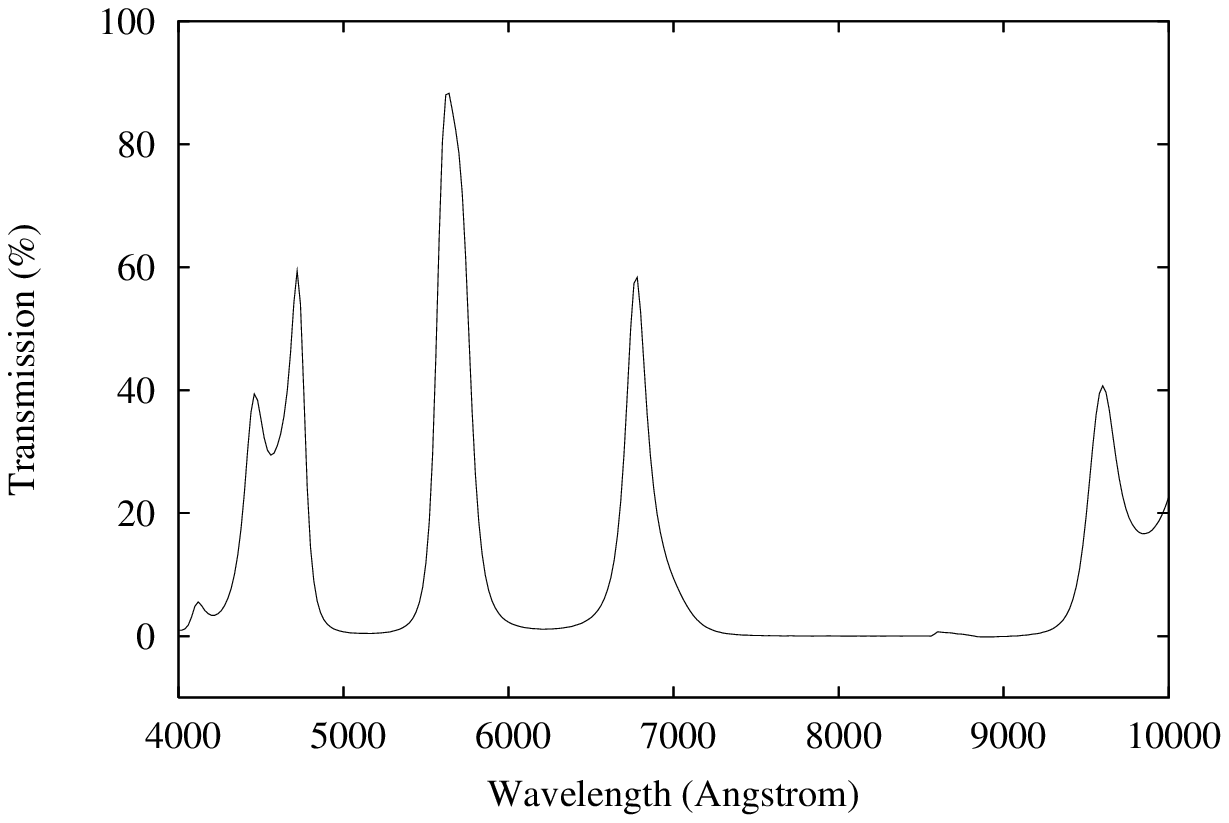}
\caption{Transmission curve of the multi-peak interference filter.}
\label{F1}
\end{figure}

The multi-peak interference filter is new in blazar monitoring and made
by Changchun Institute of Optics, Fine Mechanics, and Physics, Chinese
Academy of Sciences. The transmission curve of the filter is shown in
Figure~\ref{F1}.  The left three passbands have narrower FWHMs than but
similar central wavelengths to the traditional broad $B$, $V$, and $R$ bands.
Here we designate them as $B'$, $V'$, and $R'$, respectively. There is still
another passband beyond 9400 {\AA}. Because the CCD has a very low response
at that wavelengths, data taken in this passband are of low quality and may
only useful for bright sources. Table~1 gives the central wavelengths,
maximum transmission, and bandwidths of the left three passbands.

\subsection{The Observations}

\begin{figure}
\plottwo{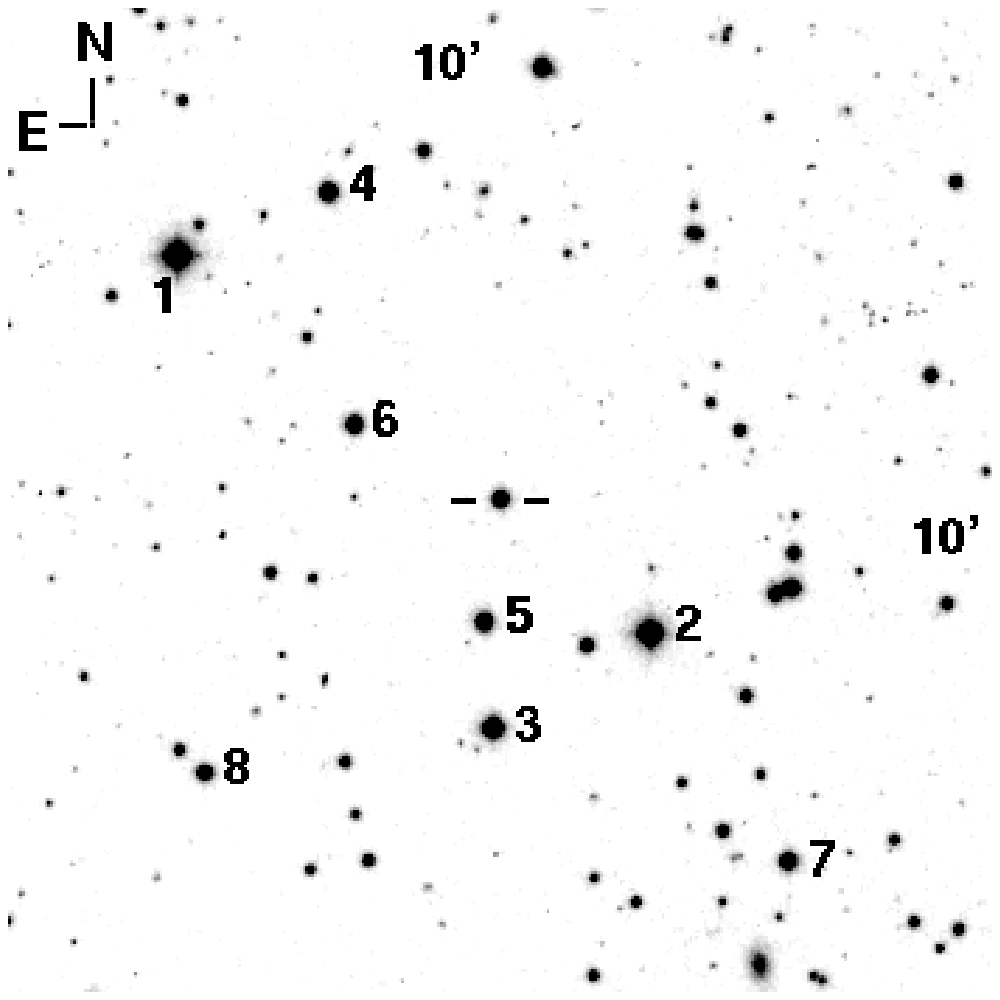}{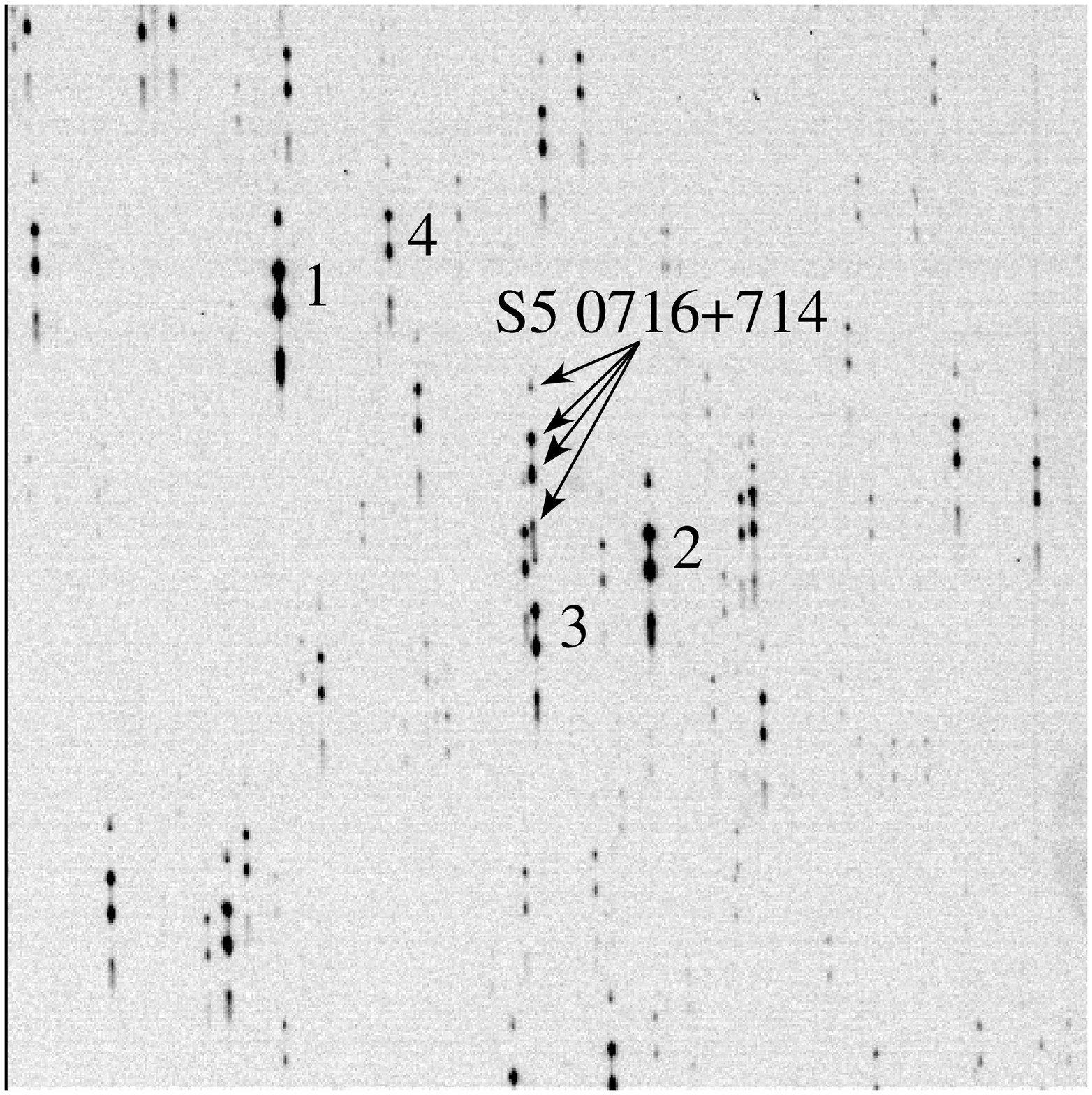}
\caption{Finding chart of S5~0716+714 (left) and example CCD frame
(right) taken with the novel photometric system. The sizes are 
$10\arcmin\times10\arcmin$ and $14\farcm5\times14\farcm5$, respectively.
Labeled on the finding chart are the BL Lac object and eight reference stars.
The first four stars are used to calibrate the BL Lac. Each object has four
images on the frame, as indicated for S5~0716+714. For those faint sources,
only two central images are visible.}
\label{F2}
\end{figure}

Figure~\ref{F2} shows the finding chart of S5~0716+714 and an example CCD
frame taken with this novel photometric system. Each source now has four
``images" (actually four segments in the spectrum of the objective prism)
on the frame. The four images, from bottom to top, correspond to the four
peaks, from left to right, in Figure~\ref{F1}. Because of the lower CCD
responses and filter transmissions at the first and fourth peaks, the
first and the fourth images are much fainter than the second and third ones.
For the faint objects, only the second and third images are visible. Therefore,
in the current paper, we do not use the data in the first and fourth passbands
in the scientific analyses, but show two example light curves in the
blue band. In the near future, a new thin CCD will be mounted to the
telescope. It has much higher responses in the blue wavelengths, so we will
have simultaneous photometries in at lease three passbands then.

\begin{figure}
\plotone{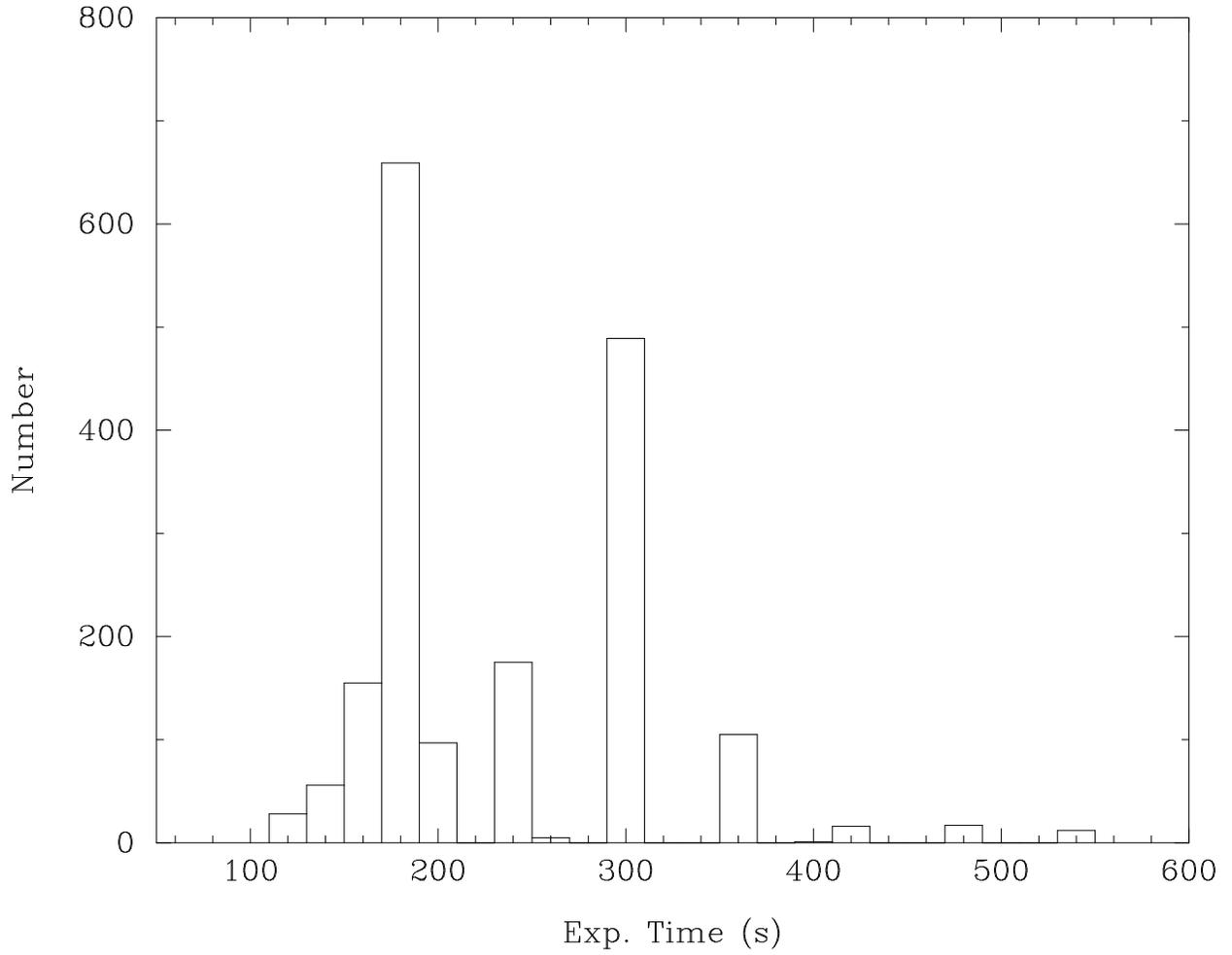}
\caption{Histogram of exposure time. Most exposure times are equal to or less
than 300 s.}
\label{F3}
\end{figure}

The monitoring of S5~0716+714 with this novel photometric system covered the
period from 2006 January 1 to February 1. As a result of weather conditions,
there are actually 19 nights' data in total. The exposure time ranges from 120
to 540 s, depending on the weather and moon phase. A histogram of the
exposure time is shown in Figure~\ref{F3}. One can see that the majority of
the exposure times do not exceed 300 s. The readout time is about 5.6 s.
So we have achieved a very high temporal resolution in multi wavebands
with this new photometric system. The observational log and parameters are
presented in Table~2.

\subsection{Data Reduction Procedures}

The procedures of data reduction include positional calibration, bias
subtraction, flat-fielding, extraction of instrumental aperture magnitude,
and flux calibration. The positional calibration was made for the $R'$-band
image, and the ``coordinates" of the $V'$- and $B'$-band images were known
by their relative positions to their corresponding $R'$-band images. The
``coordinates" in the $V'$ and $B'$ bands are, of course, not the coordinates
in the common sense but are only labels for us to find the images during the
photometry. The bias subtraction and flat-fielding were done as for the
normal CCD frames.

\begin{figure}
\plotone{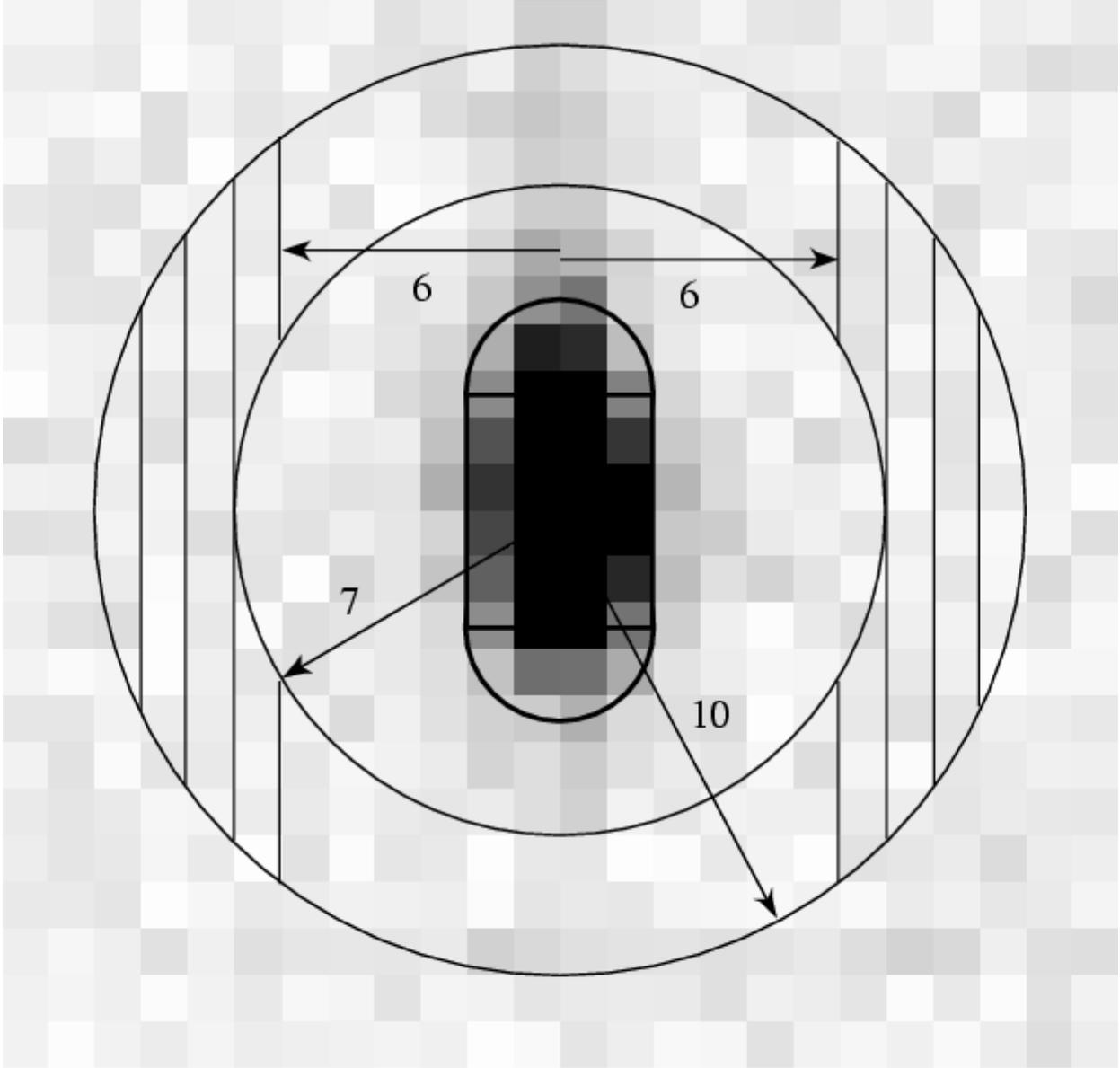}
\caption{Aperture and sky annulus definitions for an example image on a frame.
The traditional circular aperture is cut into two semi-circles along the RA
direction, and a rectangle is inset into them. An elongated aperture is thus
obtained. Images in different colors have different sizes of rectangles and
semi-circles. The shaded areas in the annulus are taken to be the sky
background. See text for details.}
\label{F4}
\end{figure}

All images are elongated in the declinational direction, especially for the
$B'$-band images. So apertures with special shapes were defined for them
and are illustrated in Figure~\ref{F4}. The traditional circular aperture
is cut into two semi-circles along the RA direction, and a rectangle
is inset into them.
An elongated aperture is thus obtained. For the $V'$- and $R'$-band images,
the widths of the rectangles (or the diameters of the semi-circles) are both
4 pixels, but the heights are 5 and 4 pixels, respectively. For the $B'$-band
image, the width and height of the rectangle are 3 and 14 pixels,
respectively. All images have a sky annulus definition with inner and outer
radii as 7 and 10 pixels, respectively. In order to eliminate the possible
contamination of the apertures to their background, areas within 6 pixels in
RA to the aperture center are excluded, and only the shaded areas are taken
to be the actual sky background.

\citet{villata98} presented eight reference stars around S5~0716+714, as
labeled in the left panel of Figure~\ref{F2}. We used the first four to
calibrate S5~0716+714.
Their broadband $B$, $V$, and $R$ magnitudes were used to mimic their
standard $B'$, $V'$, and $R'$ ones during the calibrations. The results are
presented in Table~2. The columns are observation date and time (UT), Julian
Date, exposure time, and $B'$, $V'$, and $R'$ magnitudes, their errors, and
the differential magnitudes of reference stars. The differential magnitude of
reference stars is defined as the magnitude difference (nightly mean set to 0)
between star 3 and the average of the first four. It can be taken as a measure
of the accuracy of the observations.

There are, of course, some systematic differences between our $B'$, $V'$, and
$R'$ system and the broadband $B$, $V$, and $R$ system. To quantitatively
assess the differences, we carried out photometric measurements for seven
reference stars (1, 2, 3, 4, 6, 7, 8) by using the same aperture and background
definitions and by using the same four reference stars for flux calibrations,
as exactly did for S5~0716+714. Star 5 was excluded because its $R'$ image is
partly overlapped with the $B'$ image of the BL Lac and its $B'$ image is very
close to the $V'$ and $R'$ images of star 3\footnote{This is another source of
error in the $B'$ band data of S5~0716+714. Star 3 was not excluded just
because its $V'$ and $R'$ images are much brighter than the $B'$ image of star
5 and thus are much less polluted.}. The photometric measurements of the seven
stars were carried out for 20 best-observed frames taken on the photometric
night of JD~2,453,742. Then the average $B'$, $V'$, and $R'$ magnitudes
were calculated for the seven stars and compared with their $B$, $V$, and $R$
magnitudes. The standard deviations of 7 $(B-B')$'s, 7 $(V-V')$'s, and 7
$(R-R')$'s were computed, and the results are 0.048, 0.028, and 0.036
mags, respectively. This indicates that there are only small systematic
difference between our $B'$, $V'$, and $R'$ system and the broadband $B$,
$V$, $R$ system.

\section{RESULTS}

\subsection{Light Curves}

\begin{figure}
\plotone{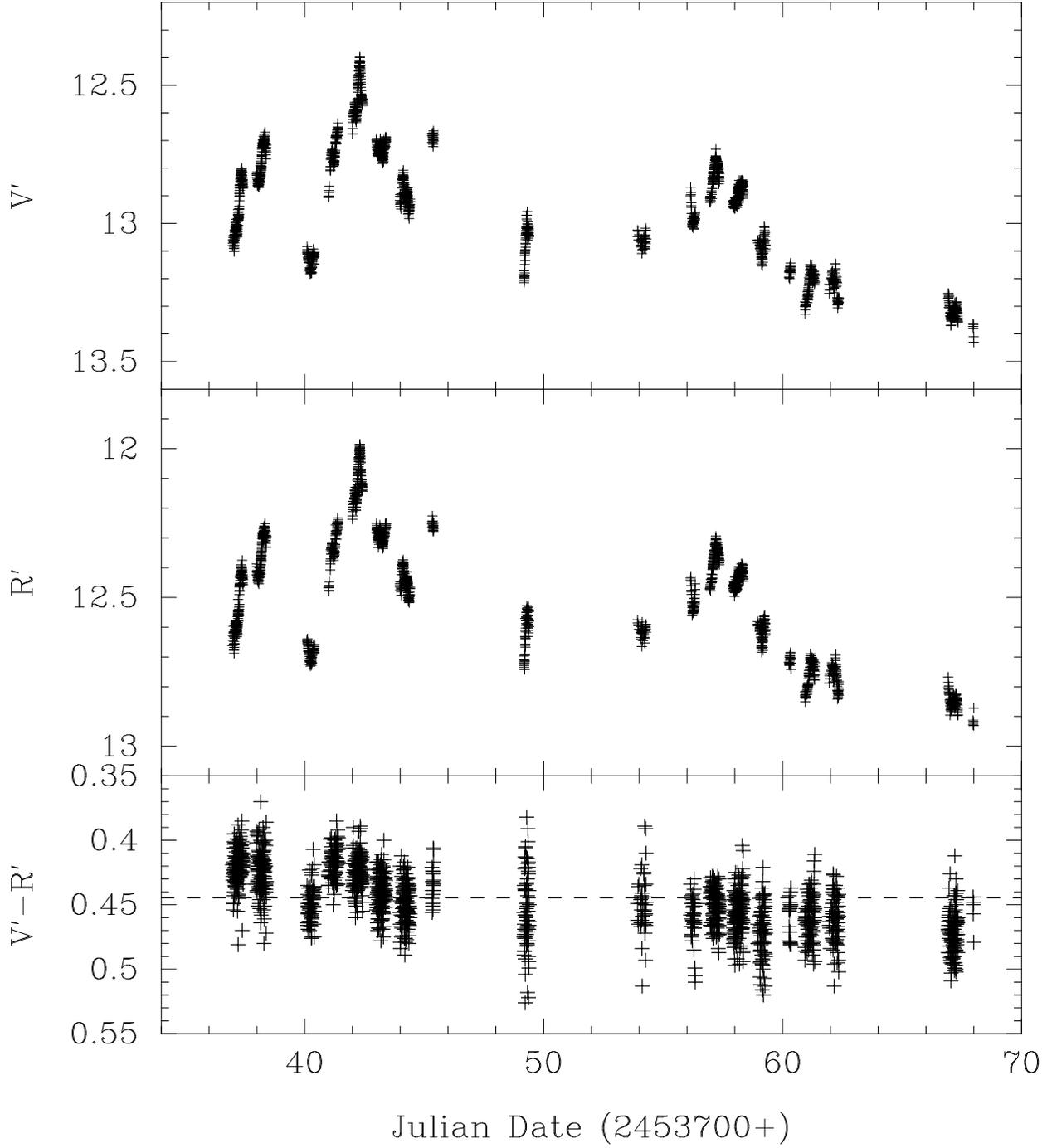}
\caption{Light curves in the $V'$ and $R'$ bands and color evolution. The
first halves of the light curves have much larger amplitudes in both
internight and intranight variations than the second halves. The colors
in the first half is much bluer than those in the second, as manifested by
the dashd line, the mean color.}
\label{F5}
\end{figure}

The $V'$- and $R'$-band light curves of the whole monitoring period are shown
in Figure~\ref{F5}. The object was very active during
this period. Two strong flares can be seen at JD~2,453,742 and 2,453,757
with amplitudes of about 0.8 and 0.6 mags, respectively. Additional flares
might also occur at JD~2,453,738, 2,453,745, and/or 2,453,749. The latter
three are less certain because of the lack of observations on one or both
sides of them. In addition to the strong internight variations, one can see
clear intranight variations as well. Except for those less sampling nights
(JD~2,453,745, 2,453,760, and 2,453,768), most intranight amplitudes are
larger than 0.1 mags and can be as large as 0.3 mags.

The whole light curve can be divided into two halves. Statistically, the
first half has larger amplitudes in both internight and intranight variations
than the second half has. So the object was more active in the first half
than in the second.

\begin{figure}
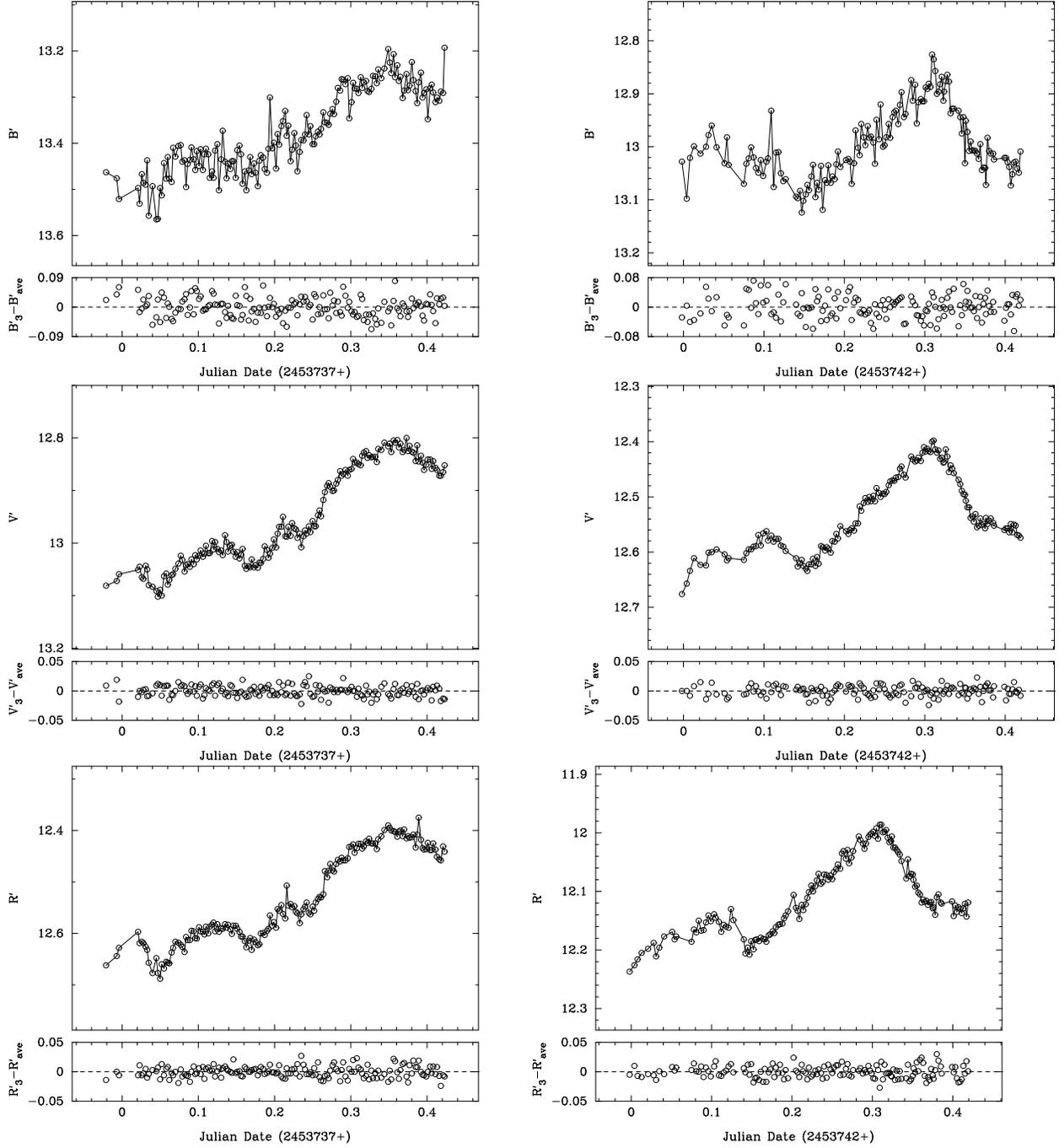

\plottwo{f6a.ps}{f6b.ps}
\plottwo{f6c.ps}{f6d.ps}
\plottwo{f6e.ps}{f6f.ps}
\caption{Intranight light curves on JD~2,453,737 and 2,453,742, in
$B'$, $V'$, and $R'$ bands from top to bottom. The large panels show the
light curves of S5~0716+714, while the small ones display the differential
magnitudes between star 3 and the average of the first four. The errors are
not plotted for clarity. The $B'$ band light curves have large errors, as
can be seen in the small panels.}
\label{F6}
\end{figure}

Figure~\ref{F6} gives the intranight light curves on JD~2,453,737
and 2,453,742, which are respectively the first and the brightest nights.
From top to bottom, they are in $B'$, $V'$, and $R'$ bands, respectively.
The large panels are the light curves of S5~0716+714, while the small ones
give the differential magnitudes (nightly mean set to 0) between star 3 and
the average of the first four.

On JD~2,453,737, the object kept on brightening, with some oscillations,
then got fainter at the end. There was a sharp increase in brightness
around JD~2,453,737.26. The $V'$ magnitude increased from 12.949 at
JD~2,453,737.26147 to 12.886 at JD~2,453,737.27124 (see Table~2), resulting
in a brightening rate of about 0.004 $\rm{mag\,min^{-1}}$, which is twice as
fast as the rate reported by \citet{villata00}. On JD~2,453,742, the object
underwent some small-amplitude oscillations at first. Then a flare was
observed with an amplitude of about 0.235 mags in the $V'$ band and a
duration of about 0.2 days. The flare peaks at JD~2,453,742.31079. It is
also the peak of the first strong flare that lasted for five days (see
Fig.~\ref{F5}).

As can be seen, our temporal resolution is very high. The typical sampling
intervals are $3.5\sim4.0$ minutes in these two nights. On both
nights, the light curves in the three bands are consistent excellently with
each other, and the differential magnitudes in the $V'$
and $R'$ bands have small rms's (0.0089 and 0.0093 for JD~2,453,737, and
0.0088 and 0.0103 for JD~2,453,742), which both signify the accuracy of our
observations. As expected, the $B'$-band data have large errors (for clarity,
the error bars are not plotted), and the curves show frequent, irregular
jumps in both large and small panels. So we did not show the $B'$ band light
curve in Figure~\ref{F5}, and will not use the $B'$ band data in the
following analyses.

\subsection{Correlations and Time Lags}

\begin{figure}
\plottwo{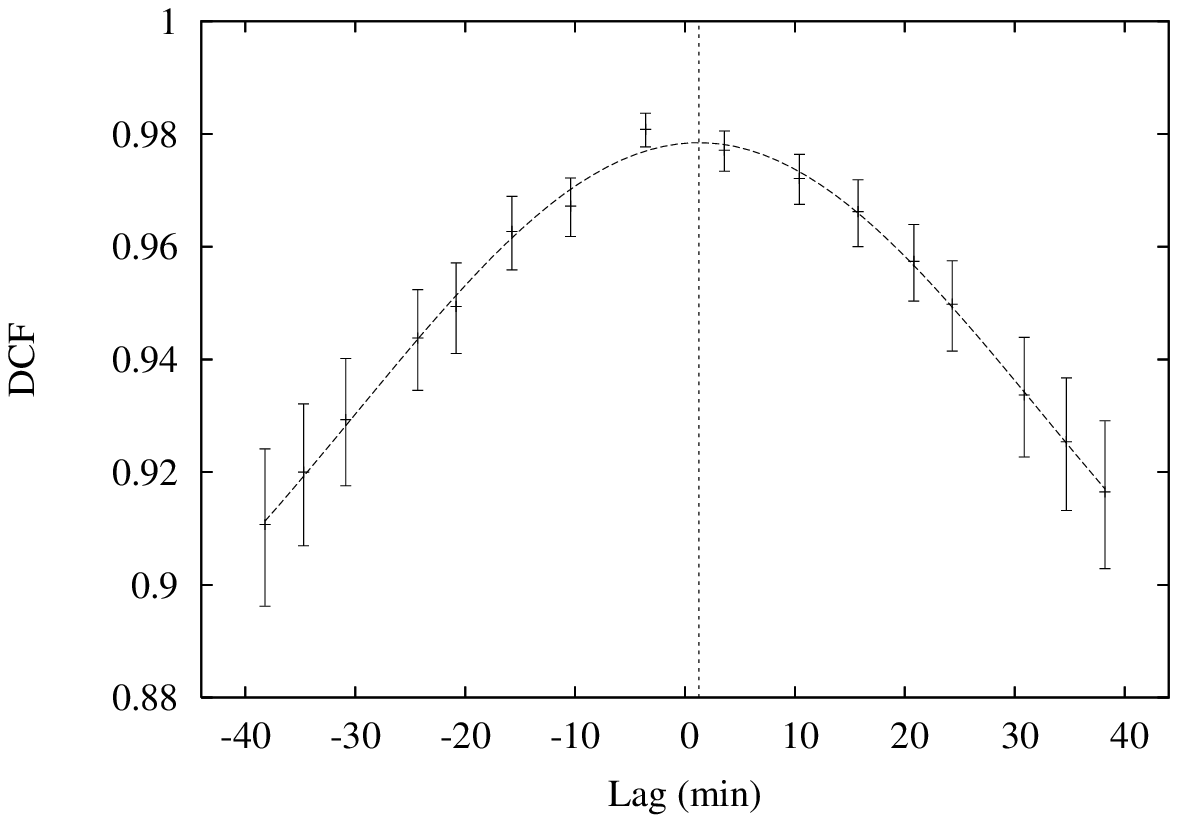}{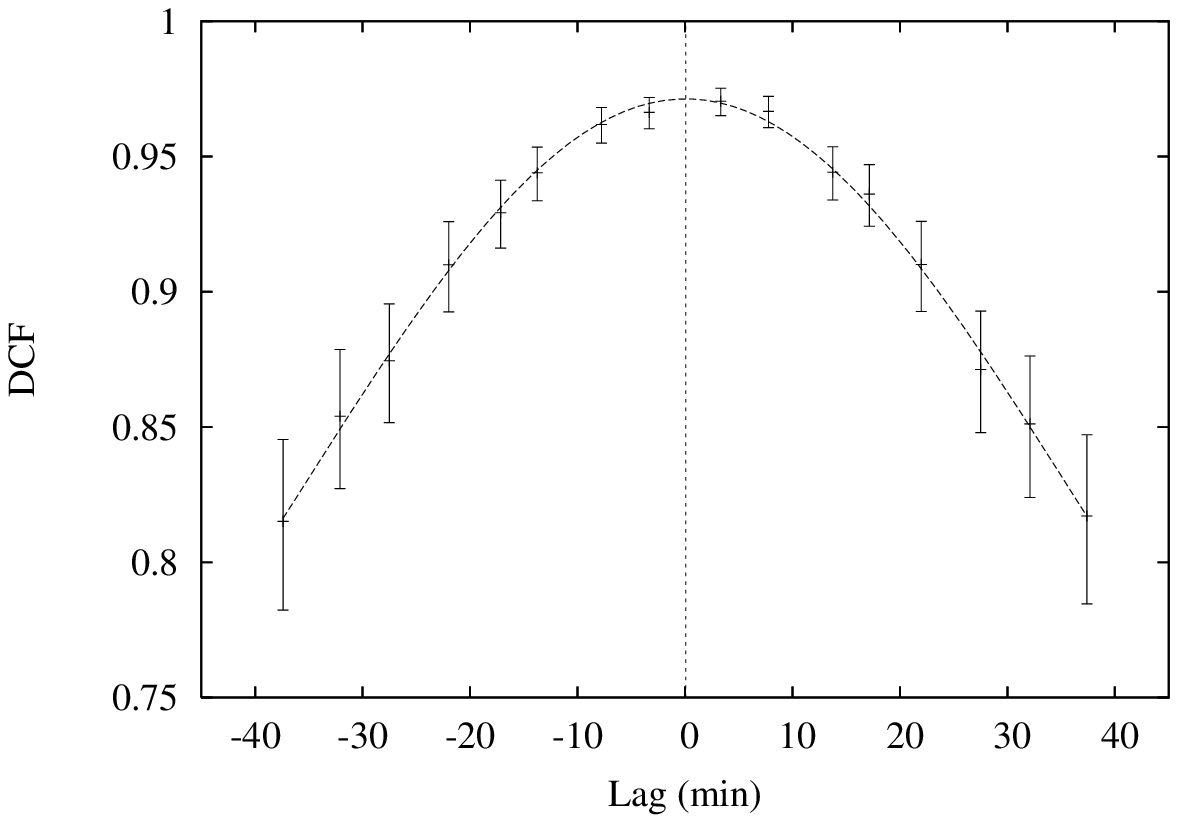}
\caption{ZDCFs between the $V'$- and $R'$-band variations on JD~2,453,737
and 2,453,742. The dashed lines are Gaussian fits to the points, and the
dotted lines (at 1.27 and 0.06 minutes for left and right, respectively) label
the centers of the Gaussian profiles, which signify the lags of the
$R'$ band to $V'$ band variations.}
\label{F7}
\end{figure}

In order to investigate the correlation between the variations in different
wave bands and to derive the possible lags between them, we at first carried
out analyses of the z-transformed discrete correlation function \citep[ZDCF;]
[]{alex97} on the $V'$- and $R'$-band variations. The ZDCFs between the $V'$-
and $R'$-band variations are presented in Figure~\ref{F7} for JD~2,453,737 and
2,453,742. The dashed lines are Gaussian fits to the points, and the dotted
lines label the centers of the Gaussian profiles, which signify the time lags
of the $R'$-band to the $V'$-band variations. The lags are $1.27\pm0.22$ and
$0.06\pm0.18$ minutes for JD~2,453,737 and 2,453,742, respectively.

At the same time, the interpolated cross-correlation function \citep[ICCF;][]
{gaskell87} was used to measure the time lags and the errors. The error
was estimated with a model-independent Monte Carlo method, and the lag was
taken as the centroid of the cross-correlation functions that were
obtained with a large number of independent Monte Carlo realizations
\citep{white94,peterson98,peterson04}. Four thousand realizations were
performed for both nights and gave the lags as $2.34\pm5.27$ on JD~2,453,737
and $-0.01\pm1.88$ minutes on JD~2,453,742.

All time lags are very short, shorter than our typical sampling interval,
and are associated with large errors. So we conclude that our observations
do not detect apparent time lag between the $V'$- and $R'$-band variations.

In optical regimes, variations at long wavelengths are usually reported to
lag those at short wavelengths. For S5~0716+714,
\citet{qian00} determined an upper limit of lag of about 6 minutes between
the $V$ and $I$ band variations. Similarly, \citet{villata00} claimed an
upper limit of 10 minutes to the possible delay between the $B$ and $I$ band
variations using high quality data. Most recently, \citet{stalin06} reported
time lags of about 6 and 13 minutes for the $V$ and $R$ band variations on two
individual nights. But they also noted that their measurement intervals were
close to these putative lags, and the lags should be treated with caution.
For other objects, for example, \citet{romero00} detected time lags of a few
to 17.3 minutes between the $V$ and $R$ band variations for PKS~0537$-$441.
However, the monitoring on 3C~66A does not strongly support any time lag
between variations at different wavelengths. The detailed spectral modeling
with a time-dependent leptonic one-zone jet model also does not predict such
spectral hysteresis (M. B\"ottcher, private communication). Our data,
characterized by very high temporal resolution and simultaneous measurements
in multi frequencies, do not detect apparent lag between variations in the
$V'$ and $R'$ bands. Compared to the much longer lags between variations in
X-rays \citep[e.g.,][$\sim1$ hr]{takahashi96} or between X-ray and EUV/UV
variations \citep[e.g.,][$1\sim2$ days]{urry97}, these very short or even
absent time lags in optical regimes may be the result of very small frequency
intervals, and may indicate that the photons in these wavelengths should be
produced by the same physical process and emitted from the same spatial region.

\subsection{Color Behavior}

\subsubsection{The Color-Magnitude Correlation}

\begin{figure}
\plottwo{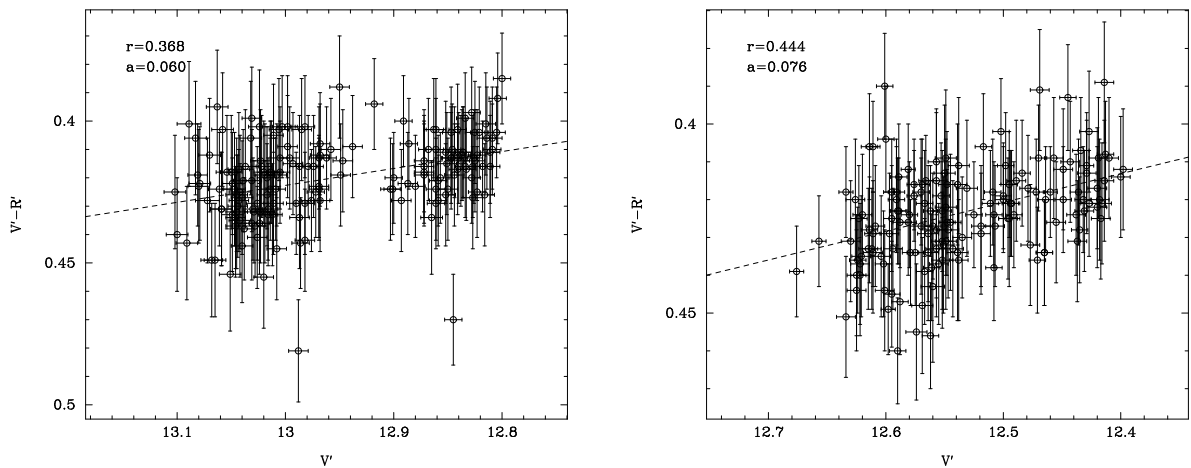}{f8b.ps}
\caption{Color versus magnitude distribution for JD~2,453,737 and
2,453,742. The dashed lines are linear fits to the points. The Pearson
correlation coefficients of 0.368 and 0.444 indicate strong correlations
between color and magnitude. The slopes of the two fits are 0.060 and 0.076,
respectively.}
\label{F8}
\end{figure}

As mentioned in \S1, the color behavior during the variations can put strong
constraints on the variation mechanisms. Figure~\ref{F8} displays the color
versus magnitude distributions on JD~2,453,737 and 2,453,742. There is a
clear bluer-when-brighter (BWB) chromatism on both nights. The dashed
lines are linear fits to the points. The Pearson correlation coefficients
are 0.368 and 0.444, and the significance levels are $1.1\times10^{-6}$ and
$1.6\times10^{-8}$, respectively, which indicate strong correlations between
color and magnitude. The dashed lines have slopes of 0.060 and 0.076 for
JD~2,453,737 and 2,453,742, respectively.

\begin{figure}
\plotone{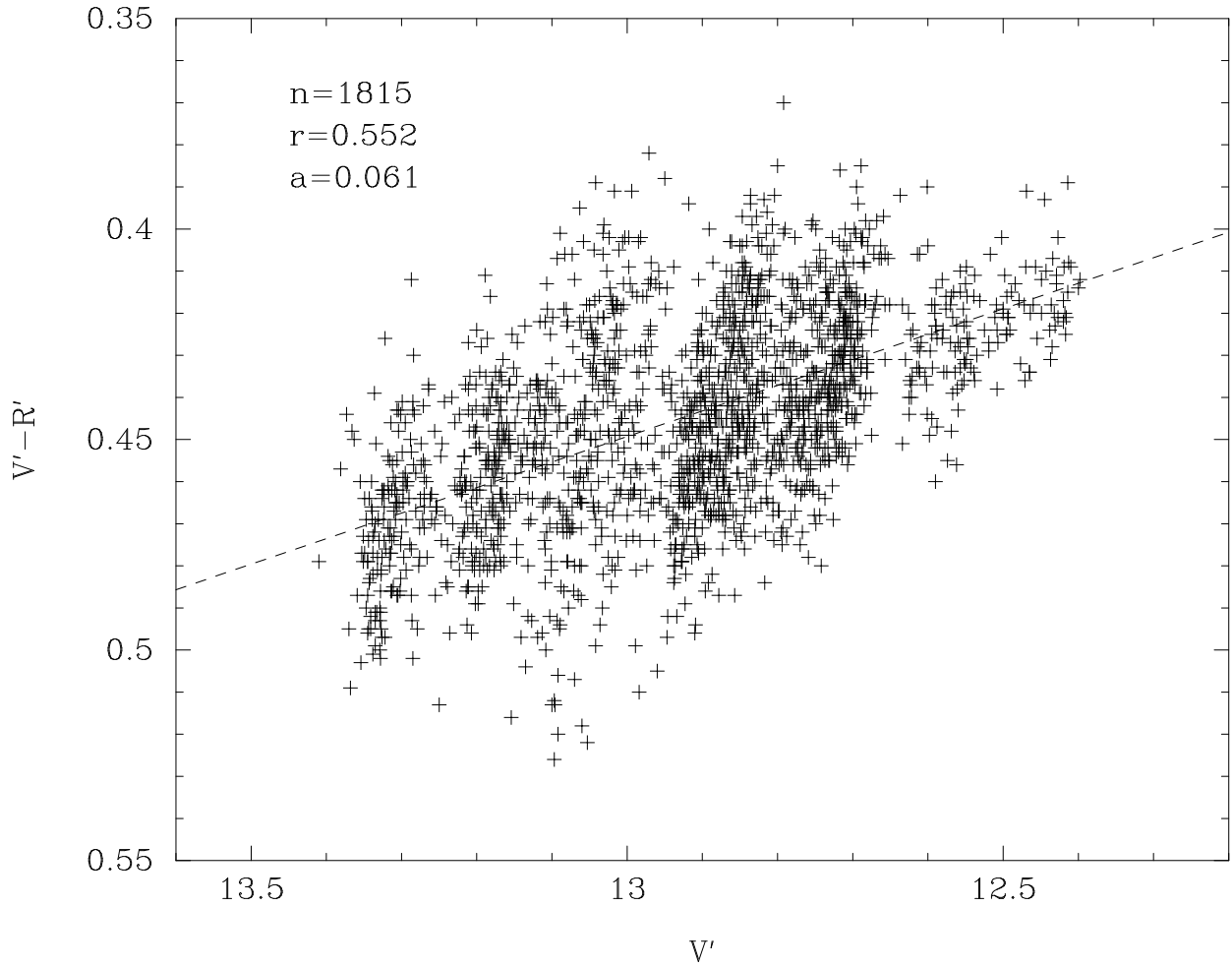}
\caption{Color versus magnitude distribution for all nights. The linear fit
to the points, the dashed line, gives a slope of 0.061. The Pearson
correlation coefficient of 0.552 means a strong correlation between color
and magnitude.}
\label{F9}
\end{figure}

For all nights, there is also a BWB trend for S5~0716+714. This is illustrated
in Figure~\ref{F9}. The dashed line is the linear fit to the 1815 points, and
it has a slope of 0.061. The Pearson correlation coefficient is 0.552,
indicating a strong correlation between the color and magnitude.

The color or spectral behaviors of S5~0716+714 and other blazars have been
investigated in optical bands by many authors \citep[][and references
therein]{vagnetti03,stalin06,wu05}. On short timescales, the blazars usually
BWB chromatism when they are in an active or
flaring state. On long timescales, an achromatic trend is usually found. Our
intranight color behavior is in agreement with previous results, whereas the
long-term internight variations show a different color behavior. Our BWB
chromatisms argue strongly against the microlensing or geometrical effects as
the variation mechanisms in S5~0716+714.

In the variability of blazars, two factors may result in color change. One
is the difference in variation steps at different frequencies, and the other,
the difference in variation amplitudes of these variations. If the variation
at high frequency leads that at low frequency, we will observe a BWB
chromatism. On the other hand, if the fluxes at different
frequencies vary simultaneously, but the amplitude is larger at high
frequency than at low one, we will again observe a BWB chromatism. In
the actual cases, it may be a combination of these two factors that results
in the color changes of blazars. In \S3.2, we do not find apparent time lag
between variations in different wave bands. The observed BWB phenomena
should be mainly attributed to the larger amplitudes at higher frequency
This will be confirmed in the next section.

\subsubsection{Color-Activity Correlation}

The bottom panel of Figure~\ref{F5} displays the temporal evolution of the
color $V'-R'$ or the $V'-R'$ ``color curve". Most nights have a scatter less
than 0.1 in color, except for JD~2,453,749 and 2,453,754. The two nights are
close to or within the phase of full moon, so the photometries have relatively
larger errors, and the colors are much more scattered than those in other
nights.

From the color curve, one can see that the object generally showed a bluer
color when it was brighter, which is consistent with the result in
Figure~\ref{F9}. However, it is also clear that the color does not correlate
solely with the brightness. For example, the object was brightest but not
bluest on JD~2,453,742. The brightness on JD~2,453,737, 2,453,738, and
2,453,740 is much fainter than that on JD~2,453,742, but the former nights
have comparable or even bluer colors than the latter night has. In analogy to
the division of the light curve, the color curve can also be divided into
two halves, i.e., those before and after JD~2,453,750. The first half is much
bluer than the second, as manifested by the dashed line, the mean color, in
the color curve in the bottom panel of Figure~\ref{F5}. Therefore, the object
seems show a bluer color when it is more active.

To study the relation between the color and the activity in more detail, we
at first need to define the activity of a blazar, or how a blazar is called
active. The activity has two aspects; one is the frequency or rate, and the
other, the amplitude. The frequency of activity can be described by duty cycle,
which is defined as the fraction of time a source spends more than $3\sigma$
away from its weekly average \citep[][]{wagner95}.

For the amplitude of activity, we define it as the sum of the intranight and
internight amplitudes
\begin{equation}
A=A_{\rm intran}+A_{\rm intern}.
\end{equation}
The intranight amplitude can be simply calculated as the difference between
the maximum and minimum magnitudes (or fluxes) of a certain night, as did in
\S3.1.3. A more precise way is to take the measurement error into account.
As in \citet{heidt96}, the intranight amplitude is defined as
\begin{equation}
A_{\rm intran}=\sqrt{(m_{\rm max}-m_{\rm min})^2-2\sigma^2},
\end{equation}
where $m_{\rm max}$ and $m_{\rm min}$ are respectively the maximum and minimum
magnitudes (or fluxes) of a certain night, and $\sigma$ is the measurement
error. Here we take $\sigma$ as the nightly average of the differential
magnitude of reference stars ($\delta V'$ or $\delta R'$ in Table~2).

For the internight amplitude, the average magnitudes (or fluxes) are
calculated at first for all nights. Then the internight amplitude of the
$i$th night is defined as
\begin{equation}
A_{{\rm intern},i}=\frac{|\langle m\rangle_i-\langle m\rangle_{i-1}|+|\langle m\rangle_i-\langle m\rangle_{i+1}|}{2},
\end{equation}
where $\langle m\rangle_{i-1}$, $\langle m\rangle_i$, and $\langle
m\rangle_{i+1}$ are the average magnitudes of the $(i-1)$th, $i$th, and
$(i+1)$th nights. When there is no observation on the $(i-1)$ or $(i+1)$
night, the data on the night closest to the $(i-1)$ or $(i+1)$ night are
used if the time differences are not large, but the resulting amplitude may
have large error. For the first and the last night
of the sequence, only the first or second absolute value is calculated in
Equation 3 and it is not divided by 2. In this paper, we will focus on the
amplitude rather than the frequency of activity. So we call a blazar
``active" when it has a large amplitude of activity.

\begin{figure}
\plotone{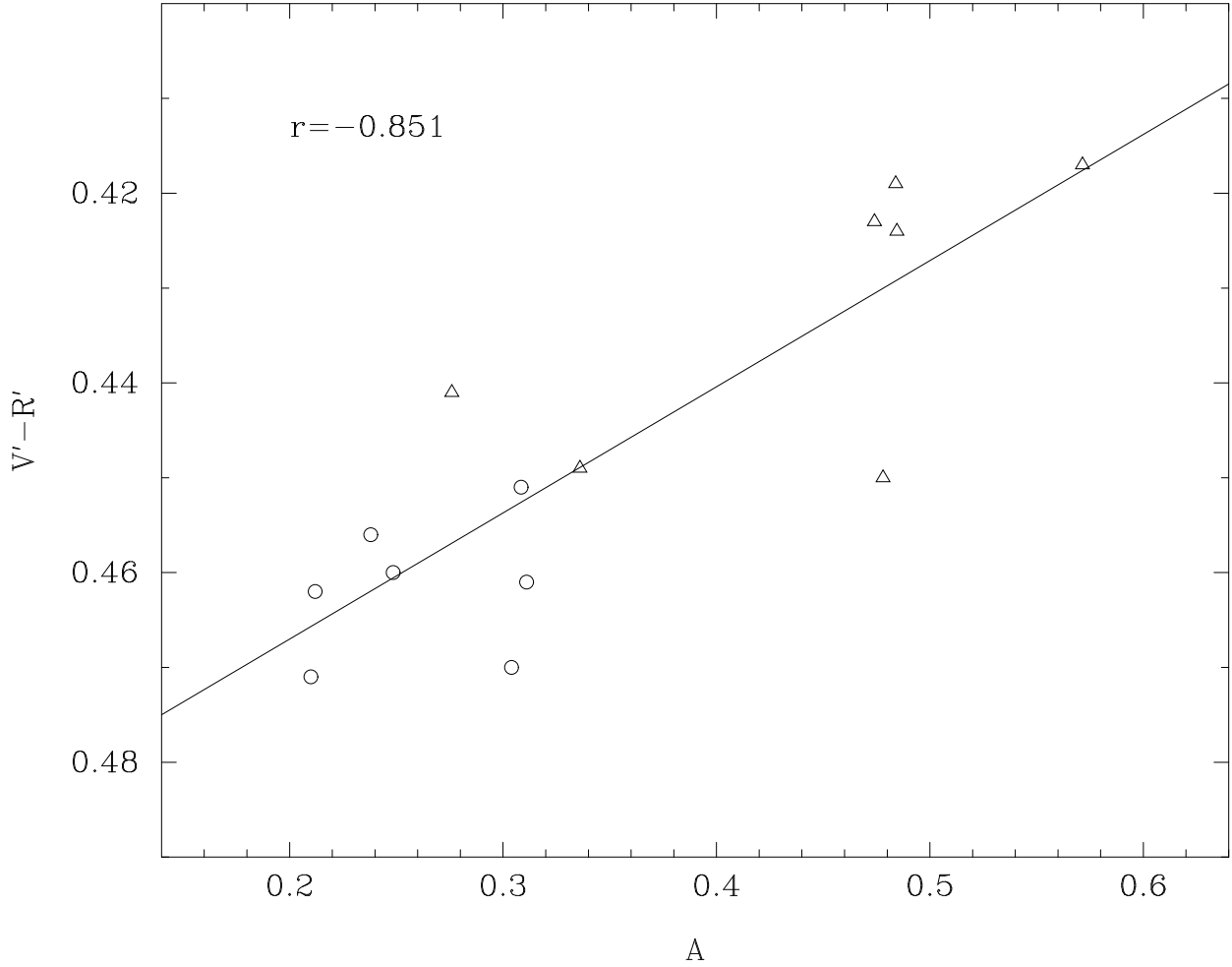}
\caption{Correlation between color and amplitude of activity. The triangles
are nights of the first half of our monitoring period, whereas the open
circles are nights of the second half. The solid line is the best fit to all
14 points. The Pearson correlation coefficients, $r=-0.851$, indicate a strong
trend that the object become bluer when it is more active.}
\label{F10}
\end{figure}

With these definitions, we calculated the intranight and internight amplitudes
of activity of S5~0716+714
on the 19 monitoring nights and listed the results in Table~3. The columns
are Julian Date, average $V'$ magnitude, internight and intranight amplitudes
in the $V'$ band, average $R'$ magnitude, internight and intranight amplitudes
in the $R'$ band, and time duration
of the monitoring in that night. Figure~\ref{F10} illustrates the relation
between the color and the amplitude of activity. We do not plot the data on
JD~2,453,749 and 2,453,754 because they have much larger measurement errors
and much larger scatters in color than those on other nights. Data on
JD~2,453,745, 2,453,760, and 2,453,768 are also excluded from the figure
because they are much less sampled than those on other nights. So there are
14 nights' data left. The triangles and circles designate nights of the first
and second halves of the light curve, respectively. The former has much bluer
color than the latter, as expected.
The solid line is the best fit to the 14 points. The Pearson correlation
coefficient is $-0.851$ and the significance level is $1.15\times10^{-4}$,
which mean strong correlation between color and amplitude of activity.
The correlation confirms the result of the above visual inspection: the object
showed a bluer color when it was more active.

\begin{figure}
\plottwo{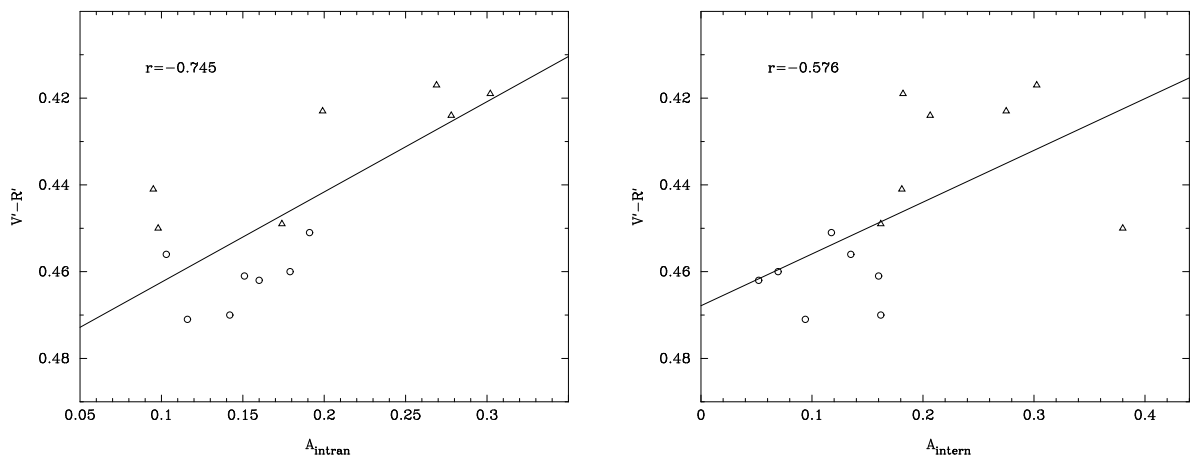}{f11b.ps}
\caption{Correlations between color and intranight (left)
and internight (right) amplitudes. The former correlation is stronger
than the latter, but both are much weaker than the correlation in
Fig.~\ref{F10}.}
\label{F11}
\end{figure}

Figure~\ref{F11} displays whether the intranight and internight amplitudes 
separately correlate with the color. The distribution on the
left panel has a Pearson correlation coefficient of $-0.745$ and a
significance level of $2.24\times10^{-3}$, while the right panel shows a
correlation coefficient of $-0.576$ and a significance level of 0.03. The
former correlation is stronger than the latter, but they are both much weaker
than the correlation in Figure~\ref{F10}.

Most observations detected a BWB chromatism in blazars. At the same time, a
minority of monitorings found a redder-when-brighter trend in some blazars
\citep[e.g.,][]{raiteri03,gu06}. This inconsistency may be explained with the
bluer-when-more-active (BWMA) chromatism: even though the object is in a
faint state, it can still have relatively large amplitude of variations, so
it can show a relatively blue color.

\section{COLOR EVOLUTION DURING THE FLARE}

According to theoretical predictions, the color of a blazar usually evolves
with a wave-like pattern during a flare, and there is usually a loop path on
the color versus magnitude (or flux) diagram, in either clockwise or
anti-clockwise direction. The direction depends on the frequency
that the observation is made and the peak frequency of the synchrotron
component in the spectral energy distribution (SED) of the blazar
\citep[see Figs. 3 and 4 in][]{kirk98}. The loop path has been frequently
observed in X-ray \citep[e.g.,][]{sembay93,takahashi96,kataoka00,zhang99,
zhang02,malizia00,ravasio04}, and, in one case, in infrared \citep{gear86}.
In optical regimes, only \citet{xilouris06} have reported a similar pattern.

\begin{figure}
\plottwo{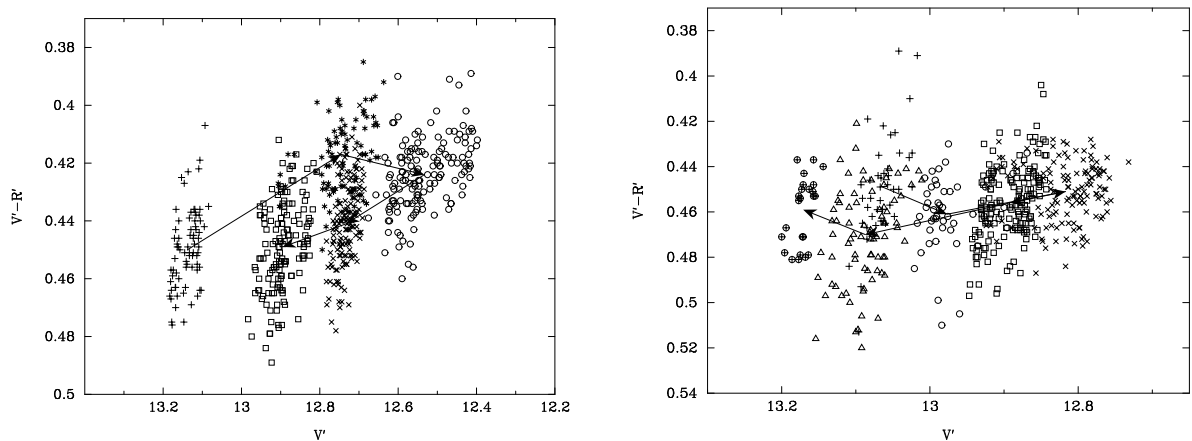}{f12b.ps}
\caption{Color versus magnitude distributions during two flares. Different
symbols signify measurements on different nights. The individual measurements
do not describe a loop path, while the nightly means define two loop paths in
clockwise direction, as indicated by the arrows.}
\label{F12}
\end{figure}

In our monitoring, we recorded two strong flares, i.e., the one from
JD~2,453,740 to 2,453,744 and the other from JD~2,453,754 to 2,453,760 (see
Fig.~\ref{F5}). The temporal evolutions of the color do follow a wave-like
pattern during the two flares (see the bottom panel of Fig.~\ref{F5}). On
the color versus magnitude diagrams in
Figure~\ref{F12}, the results are controversial. When the individual
measurements are considered, both diagrams are just messy; when we focus only
on the nightly means, there is a loop in clockwise direction on both diagrams,
as labeled by the arrows.

The loop path can be explained by a spectral hysteresis, i.e., the variation
in one band lags that in the other. Our observations do not find apparent
time lags between the intranight variations in different wave bands, so the
individual measurements do not show a loop path. However, the loop paths
described by the nightly means may suggest that S5~0716+714 has a spectral
hysteresis in its long-term (from hours to days) variations. By using the same
Monte Carlo algorithm of \citet{white94} and \citet{peterson98,peterson04},
we derived a time lag of $0.10\pm3.98$ days between the $V'$- and $R'$-band
nightly-mean light curves. Given the large error and the short lag relative
to the sampling interval (1 day), the result is of relatively low significance
level in quantity, but should be qualitatively reasonable.

\section{CONCLUSIONS AND DISCUSSIONS}

In this paper, we have introduced a novel photometric system, which consists
of a Schmidt telescope, an objective prism, a CCD camera, and, especially a
multi-peak interference filter. The multi-peak interference filter enables
light in multi passbands to pass through it simultaneously. The light in
different passbands are differentially refracted by the objective prism and
focus on the CCD separately, so we have multi images for each object on the
CCD frames. This system enables us to carry
out simultaneous photometric measurements in multi wavebands, and it has the
advantage of being invulnerable to the possible instability of instruments
and weather. When using for blazar monitorings, the system can accurately
trace the color change during the variation, which is crucial in constraining
the variation mechanisms of blazars.

The multi-peak interference filter is new, at least in blazar monitorings.
The transition from the traditional one-passband filter to
the new multi-peak interference filter is analogous to the transition
from the traditional slit spectrograph to the fiber spectrograph. The new
filter can greatly increase the efficiency of data sampling and is worthy
being extended to the observations of other targets, especially those
variable sources.

This novel photometric system was used in the monitoring of S5~0716+714 and
we achieved a very high temporal resolution. The BL Lac object was very
bright and very active during our monitoring period and showed strong
variations on both internight and intranight timescales. Two strong flares
were observed, with amplitudes of variations of about 0.8 and 0.6 mags,
respectively. There may be three more but less certain flares. Significant
BWB correlations were found for both internight and intranight variations,
and the BWB internight variability is different from the basically achromatic
long-term variability of S5~0716+714 observed previously \citep{ghise97,
raiteri03}. No apparent time lag was found between the $V'$- and $R'$-band
variations. The BWB chromatism should be mainly attributed to the larger
amplitude of variation at higher frequency.

In addition to the BWB chromatism, we also found a BWMA chromatism, which is
new and has not been reported before. It helps to explain some exceptional
color behaviors (e.g., redder-when-brighter) observed in blazars sometimes.
The theoretical reason for the BWMA chromatism is still unknown.
Phenomenologically, when the object is more active (has larger variation
amplitude), the difference in variation amplitudes in different wavebands is
larger, so the object is bluer. Unlike the BWB chromatism, which can be either
long- or short-term behavior, the BWMA phenomenon is only a long- or, at least,
intermediate-term behavior. To reveal or confirm the BWMA chromatism, the
monitoring should cover a relatively long period and consists of as many
successive nights as possible. Intensive intranight observations are also
desired.

The color behaviors can put strong constraints to the variation mechanisms
of blazars. The mechanisms can be broadly classified into extrinsic and
intrinsic origins \citep{wagner95}. The extrinsic origins include the ISS and
gravitational microlensing effects. The ISS can explain variations at low
radio frequencies, but can not produce variations at optical wavelengths. The
microlensing is an achromatic process and should result in strictly symmetric
light curves, which are both contrary to our results. So the variations of
S5~0716+714 observed by us are unlikely to have an extrinsic origin. The
intrinsic origins include the shock-in-jet scenario, accretion disk
instability, and geometrical effects. In the accretion disk scenario, external
gravitational perturbations may induce flares and hot spots in the accretion
disk around a black hole. This may account for the optical-UV microvariability
observed in blazars \citep{chak93,mangalam93}, but it has difficulty in
explaining the strong or rapid flares in blazars, as observed by us. It also
can not account for the correlated radio-optical variations observed in
S5~0716+714 \citep{quir91}. In fact, the general absence of the big blue bump
in the SED of blazars also argue against a significant contribution of the
thermal emission from the accretion disk to the overall emission of blazars.
The geometrical effects are that the helical or precessing jet leads to
varying Doppler boosting towards the observers and hence the flux variations.
They usually result in achromatic and periodic variations, which is also not
the case in our results. Now the other mechanisms are all excluded, the
shock-in-jet scenario becomes most plausible. In fact, the irregular
variations and, especially, the BWB color behaviors are all consistent with
the predictions of the shock-in-jet model.

Variation mechanism is an essential issue in the investigations of blazars.
Our new photometric system has the advantages of accurate color trace, high
temporal resolution, and being invulnerable to instrumental and weather
instabilities. A new thin CCD camera will soon be mounted to the
telescope, and it has much higher responses in blue wavelengths than the
current thick CCD has. The objective prism may also soon be replaced by
an diffraction grating. The upgraded system will be used to monitor a few
bright blazars in the near future, and the simultaneous three-band
photometries will help to investigate the variation mechanisms in them with
high confidence level.

\begin{acknowledgements}
The authors thank the anomymous referee for constructive suggestions and
insightful comments.
We thank Dr. Bradley M. Peterson for sending the code for
calculating the time lags between variations in different wave bands. This
work has been supported by the Chinese National Natural Science Foundation
grants 10603006, 10473012, 10573020, 10633020 and 10303003.
\end{acknowledgements}

\clearpage

\begin{deluxetable}{cccc}
\tablewidth{0pt}
\tablecaption{Parameters of the Left Three Passbands of the Filter}
\tablehead{\colhead{Passband} & \colhead{Wavelength} & \colhead{Max Trans} & \colhead{FWHM}\\
    & \colhead{(\AA)} & \colhead{(\%)} & \colhead{(\AA)} \\ }
\startdata
$B'$ & 4575.2 & 59.40 & 355 \\
$V'$ & 5672.9 & 88.28 & 210 \\
$R'$ & 6786.6 & 58.34 & 162 \\
\enddata
\end{deluxetable}

\begin{deluxetable}{lcccccrccrccr}
\tablewidth{0pt}
\rotate
\tabletypesize{\footnotesize}
\tablecaption{Observational log and results in the $B'$, $V'$, and $R'$ bands.}
\tablehead{\colhead{Date} & \colhead{Time} & \colhead{Julian Date} & \colhead{Exp.} & \colhead{$B'$\tablenotemark{a}} & \colhead{$B'_{\rm err}$} & \colhead{$\delta B'$} & \colhead{$V'$} & \colhead{$V'_{\rm err}$} & \colhead{$\delta V'$} & \colhead{$R'$} & \colhead{$R'_{\rm err}$} & \colhead{$\delta R'$} \\
  &  &  & \colhead{(s)} & \colhead{(mag)} & \colhead{(mag)} & \colhead{(mag)} & \colhead{(mag)} & \colhead{(mag)} & \colhead{(mag)} & \colhead{(mag)} & \colhead{(mag)} & \colhead{(mag)} 
}
\startdata
 2006 01 01 & 11:30:07 & 2453736.97925 & 240 & 13.463 & 0.027 &  0.022 & 13.081 & 0.009 &  0.009 & 12.662 & 0.009 & $-$0.014 \\
 2006 01 01 & 11:49:27 & 2453736.99268 & 180 & 13.476 & 0.027 &  0.039 & 13.072 & 0.011 &  0.019 & 12.644 & 0.012 &  0.000 \\
 2006 01 01 & 11:54:06 & 2453736.99585 & 180 & 13.521 & 0.031 &  0.061 & 13.059 & 0.010 & $-$0.018 & 12.628 & 0.011 & $-$0.006 \\
 2006 01 01 & 12:30:03 & 2453737.02075 & 180 & 13.497 & 0.026 &  0.053 & 13.051 & 0.010 & $-$0.010 & 12.597 & 0.010 & $-$0.006 \\
 2006 01 01 & 12:33:27 & 2453737.02319 & 180 & 13.531 & 0.028 & $-$0.015 & 13.045 & 0.010 &  0.000 & 12.619 & 0.009 &  0.011 \\
\enddata
\tablecomments{Table 2 is published in its entirety in the electronic edition
of the Astronomical Journal. A portion is shown here for guidance regarding
its form and content.}
\tablenotetext{a}{The $B'$ band data have large errors and should be used
with great caution.}
\end{deluxetable}

\begin{deluxetable}{cccccccc}
\tablewidth{0pt}
\tabletypesize{\small}
\tablecaption{Average Magnitudes and Variation Amplitudes on Individual Nights}
\tablehead{\colhead{Julian Date} & \colhead{$\langle V'\rangle$} & \colhead{$A_{\rm V',intern}$} & \colhead{$A_{\rm V',intran}$\tablenotemark{a}} & \colhead{$\langle R'\rangle$} & \colhead{$A_{\rm R',intern}$} & \colhead{$A_{\rm R',intran}$\tablenotemark{a}} & \colhead{Duration} \\
 & \colhead{(mag)} & \colhead{(mag)} & \colhead{(mag)} & \colhead{(mag)} & \colhead{(mag)} & \colhead{(mag)} & \colhead{(day)}
}
\startdata
2453737 & 12.952 & 0.182 & 0.302 & 12.533 & 0.186 & 0.298 & 0.444 \\
2453738 & 12.770 & 0.275 & 0.199 & 12.347 & 0.263 & 0.202 & 0.403 \\
2453740 & 13.138 & 0.380 & 0.098 & 12.688 & 0.350 & 0.091 & 0.317 \\
2453741 & 12.746 & 0.302 & 0.269 & 12.329 & 0.289 & 0.246 & 0.423 \\
2453742 & 12.533 & 0.207 & 0.278 & 12.109 & 0.201 & 0.251 & 0.421 \\
2453743 & 12.733 & 0.181 & 0.095 & 12.292 & 0.168 & 0.084 & 0.411 \\
2453744 & 12.895 & 0.162 & 0.174 & 12.446 & 0.154 & 0.143 & 0.429 \\
2453745 & 12.691 & 0.204 & 0.055 & 12.257 & 0.189 & 0.049 & 0.076 \\
2453749 & 13.062 & 0.186 & 0.242 & 12.606 & 0.179 & 0.212 & 0.223 \\
2453754 & 13.063 & 0.044 & 0.077 & 12.614 & 0.053 & 0.075 & 0.359 \\
2453756 & 12.977 & 0.160 & 0.151 & 12.516 & 0.150 & 0.131 & 0.196 \\
2453757 & 12.817 & 0.117 & 0.191 & 12.366 & 0.110 & 0.184 & 0.402 \\
2453758 & 12.892 & 0.135 & 0.103 & 12.436 & 0.125 & 0.096 & 0.420 \\
2453759 & 13.087 & 0.162 & 0.142 & 12.617 & 0.160 & 0.122 & 0.334 \\
2453760 & 13.171 & 0.065 & 0.055 & 12.712 & 0.070 & 0.055 & 0.087 \\
2453761 & 13.216 & 0.069 & 0.179 & 12.756 & 0.074 & 0.161 & 0.417 \\
2453762 & 13.226 & 0.052 & 0.160 & 12.764 & 0.047 & 0.151 & 0.405 \\
2453767 & 13.320 & 0.094 & 0.116 & 12.849 & 0.085 & 0.128 & 0.409 \\
2453768 & 13.387 & 0.067 & 0.065 & 12.915 & 0.066 & 0.046 & 0.026 \\
\enddata
\tablenotetext{a}{The spurious measurements were excluded when calculating
the intranight amplitudes.}
\end{deluxetable}

\end{document}